\documentclass[%
 reprint,
 nofootinbib,
 amsmath,amssymb,
 aps,
 floatfix,
]{revtex4-2}

\usepackage[utf8]{inputenc}
\usepackage{graphicx}
\usepackage{dcolumn}
\usepackage{bm}
\usepackage{enumitem}
\usepackage{comment}
\usepackage{siunitx}
\usepackage{booktabs}
\usepackage{hyperref}
\hypersetup{
    colorlinks = false
}
\usepackage{amsfonts}       
\usepackage{nicefrac}       
\usepackage{microtype}      
\usepackage{lipsum}
\usepackage{amsmath}
\usepackage{mathtools}
\usepackage{multirow}
\usepackage{listings} 
\usepackage{color}
\usepackage{xspace}
\usepackage{todonotes}

\usepackage{lineno}
\usepackage[normalem]{ulem}

\suppressfloats

\newcommand{\unit}[1]{\ensuremath{\text{\,#1}}\xspace}
\newcommand{\hlsfml}{\texttt{hls4ml}\xspace}
\newcommand{\Keras}{\textsc{Keras}\xspace}
\newcommand{\QKeras}{\textsc{QKeras}\xspace}
\newcommand{\Tensorflow}{\textsc{TensorFlow}\xspace}

\newcommand{\Rz}{\ensuremath{\mathrm{R}_\textit{z}}\xspace}
\newcommand{\Dkl}{\ensuremath{\mathrm{D_{KL}}}\xspace}
\newcommand{\pt}{\ensuremath{p_{\mathrm{T}}}\xspace}
\newcommand{\GeV}{\ensuremath{\,\text{Ge\hspace{-.08em}V}}\xspace}

\newcommand{\ptvecmiss}{\ensuremath{{\vec p}_{\mathrm{T}}^{\kern1pt\text{miss}}}\xspace}

\begin{document} 

\title{Autoencoders on FPGAs for real-time, unsupervised new physics detection at 40\unit{MHz} at the Large Hadron Collider}

\author{Ekaterina Govorkova}
\email{E-mail: katya.govorkova@cern.ch}
\author{Ema Puljak} 
\author{Thea Aarrestad} 
\author{Thomas James} 
\author{Vladimir Loncar}
\thanks{Also at Institute of Physics Belgrade, Serbia}
\author{Maurizio Pierini}
\author{Adrian Alan Pol} 
\author{Nicol\`{o} Ghielmetti}
\thanks{Also at Politecnico di Milano, Italy}
\author{Maksymilian Graczyk}
\thanks{Also at Imperial College London, UK}
\author{Sioni Summers}
\affiliation{European Organization for Nuclear Research (CERN)
  CH-1211 Geneva 23, Switzerland}

\author{Jennifer Ngadiuba}
\thanks{Also at California Institute of Technology, USA}
\affiliation{Fermi National Accelerator Laboratory, Batavia, IL 60510, USA}
\author{Thong Q. Nguyen}
\affiliation{California Institute of Technology
    Pasadena, CA 91125, USA}
\author{Javier Duarte}  
\affiliation{University of California San Diego
  La Jolla, CA 92093, USA }
\author{Zhenbin Wu}
\affiliation{University of Illinois at Chicago
  Chicago, IL 60607, USA
}
\date{\today}

\begin{abstract}
  In this paper, we show how to adapt and deploy anomaly detection algorithms based on deep autoencoders, for the unsupervised detection of new physics signatures in the extremely challenging environment of a real-time event selection system at the Large Hadron Collider (LHC). We demonstrate that new physics signatures can be enhanced by three orders of magnitude, while staying within the strict latency and resource constraints of a typical LHC event filtering system. This would allow for collecting datasets potentially enriched with high-purity contributions from new physics processes. Through per-layer, highly parallel implementations of network layers, support for autoencoder-specific losses on FPGAs and latent space based inference, we demonstrate that anomaly detection can be performed in as little as $80\,$ns using less than 3\% of the logic resources in the Xilinx Virtex VU9P FPGA. Opening the way to real-life applications of this idea during the next data-taking campaign of the LHC.

\end{abstract}
\maketitle

\section{Introduction}
The CERN Large Hadron Collider (LHC)~\cite{LHC} generates 40 million proton-proton collision events per second. Particles produced in these events are detected in the sensors of detectors located around the LHC ring, producing hundreds of terabytes of data per second. The largest general-purpose particle detectors at LHC, ATLAS~\cite{ATLAS} and CMS~\cite{CMS}, discard most of the collision events with online selection systems, as a result bandwidth limitations. These systems consist of two stages; the \hbox{level-1} trigger (L1T)~\cite{CMSL1T,CMSP2L1T,ATLASL1T,ATLASP2L1T}, where algorithms are deployed as programmable logic on custom electronic boards equipped with field-programmable gate arrays (FPGAs), and the High Level Trigger (HLT), where selection algorithms asynchronously process the events accepted by the L1T on commercially available CPUs. The largest fraction of events are discarded at the first selection stage, the L1T, which has the task of reducing the event rate by $2.5$ orders of magnitude within a few microseconds.
The trigger selection algorithms running in the L1T and HLT are designed to guarantee a high acceptance rate for certain physics processes under study. 
When designing searches for new physics kinds of collisions (e.g., dark matter production), physicists typically consider specific scenarios motivated by theoretical considerations. 
This \textit{supervised} strategy has proven to be successful when dealing with  theory-motivated searches, as was the case with the search for the Higgs boson~\cite{Aad:2012tfa,Chatrchyan:2012ufa}. 
Conversely, this approach may become a limiting factor in the absence of a strong theoretical prior.
For this reason, there are several community efforts to investigate unsupervised machine learning~(ML) techniques for new physics searches~\cite{aarrestad2021dark,kasieczka2021lhc}.
These investigate the use of autoencoders~(AEs) and variational autoencoders~(VAEs) for offline processing~\cite{kingma2014auto,rezende2014stochastic}, and therefore do not consider constraints such as resource usage and latency. 
Ref.~\cite{Cerri:2018anq,Knapp:2020dde} propose to integrate unsupervised learning algorithms in the online selection system of the CMS and ATLAS experiments, in order to preserve rare events which would not otherwise be selected, in a special data stream. 
A similar, albeit non-learning, approach was pursued in CMS with the \textit{exotica hotline}~\cite{symmetry-magazine, Francesco:1306501} during the first year of LHC data taking and in similar efforts during experiments at the Super Proton–Antiproton Synchrotron and Tevatron.

While the primary focus for online unsupervised learning so far has been for the HLT, this strategy could be more effective if deployed in the L1T, i.e. before any selection bias is introduced. 
Due to the extreme latency and computing resource constraints of the L1T, only relatively simple, mostly theory-motivated selection algorithms are currently deployed. 
These usually include requirements on the minimum energy of a physics object, such as a reconstructed lepton or a jet, effectively excluding lower-energy events from further processing. 
Instead, by deploying an unbiased algorithm which selects events based on their degree of abnormality, rather than on the amount of energy present in the event, we can collect data in a signal-model-independent way. 
Such an anomaly detection (AD) algorithm is required to have extremely low latency because of the restrictions imposed by the L1T.

Recent developments of the \hlsfml library allow us to consider, for the first time, the possibility of deploying an AD algorithm on the FPGAs mounted on the L1T boards. 
The \hlsfml library is an open-source software, developed to translate neural 
networks~\cite{Duarte:2018ite,bnnpaper,Iiyama:2020wap,aarrestad2021fast,Heintz:2020soy} and boosted decision trees~\cite{Summers:2020xiy} into FPGA firmware.
A fully on-chip implementation of the machine learning model is used in order to stay within the 1\,$\mu$s latency budget imposed by a typical L1T system.
Additionally, the initiation interval~(II) of the algorithm should be within 150\unit{ns}, which is related to the bunch-crossing time for the upcoming period of the LHC operations.
Since there are several L1T algorithms deployed per FPGA, each of them should use much less than the available resources.
With its interface to \QKeras~\cite{qkeras}, \hlsfml supports quantization-aware training (QAT)~\cite{AutoQ}, which makes it possible to drastically reduce the FPGA resource consumption while preserving accuracy. 
Using \hlsfml we can compress neural networks to fit the limited resources of an FPGA.

In this paper, we discuss how to adapt and improve the strategy presented in Ref.~\cite{Cerri:2018anq} to fit the L1T infrastructure. 
We focus on AEs, with specific emphasis on VAEs~\cite{kingma2014auto,rezende2014stochastic}. 
We consider both fully-connected and convolutional architectures, and discuss how to compress the model through pruning~\cite{optimalbraindamage,han2016deep,stateofpruning}, the removal of unnecessary operations, and quantization~\cite{bertmoons,NIPS2015_5647,zhang2018lq,JMLR:v18:16-456,xnornet, micikevicius2017mixed,Zhuang_2018_CVPR, wang2018training,bnnpaper}, the reduction of the precision of operations, at training time. 

As discussed in Ref.~\cite{Cerri:2018anq}, one can train (V)AE on a given data sample by minimizing a measure of the distance between the input and the output (the loss function). 
This strategy, which is very common when using (V)AEs for anomaly detection~\cite{an2015variational}, comes with practical challenges when considering a deployment on FPGAs. 
The use of high-level features is not optimal because it requires time-consuming data preprocessing. 
The situation is further complicated for VAEs, which require a random sampling from a Gaussian distribution in the latent space. 
Furthermore, one has to buffer the input data on chip while the output is generated by the FPGA processing in order to compute the distance afterwards. 
To deal with all of these aspects, we explore different approaches and compare the accuracy, latency and resource consumption of the various methods. 

In addition, we  discuss how to customize the model compression in order to better accommodate for unsupervised learning. 
Previously, we showed that QAT can result in a large reduction in resource consumption with minor accuracy loss for supervised algorithms~\cite{AutoQ,aarrestad2021fast}. 
In this paper, we extend and adapt that compression workflow to deal with the specific challenge of compressing autoencoders used for AD. 
Several approaches are possible:
\begin{itemize}
\item Post-training quantization (PTQ)~\cite{Duarte:2018ite,nagel2019datafree,han2016deep,meller2019same,zhao2019improving,banner2019posttraining}, consisting of applying a fixed-point precision to a floating-point baseline model. 
This is the simplest quantization approach, typically resulting in good algorithm stability, at the cost of losing performance. 
More aggressive PTQ (lower precision) is usually accompanied by a larger reduction in accuracy. 
\item QAT, consisting of imposing the fixed-point precision constraint at training time, e.g., using the \QKeras library. 
This approach typically allows one to limit the accuracy loss when imposing a higher level of quantization, finding a better weight configuration than what one can get with PTQ. 
However, applying QAT to VAE models for AD can result in unstable performance because QAT would return the best input-to-output reconstruction performance, but the best reconstruction does not necessarily guarantee the best AD performance. 
Ultimately, the stability of the result depends on the nature of the detected anomaly.
\item Knowledge distillation with QAT: one could change the quantized-model optimization strategy, reframing the problem as knowledge distillation~\cite{9195219,polino2018model,gao2019embarrassingly,mishra2018apprentice}. 
Rather than fixing the quantized weights to minimize the VAE loss, the difference between the loss from the quantized model and the floating-point model for the same input could be minimized.
Rather than training a quantized copy of a given floating-point model, one could train a different model to predict this floating-point output, starting from the same input. 
Doing so, one could aim at targeting the floating-point AD performance with a completely different network (e.g., an MLP regression) that could better meet the constraints of a L1T environment, e.g. being faster or consuming less computing resources.
\item Anomaly classification with QAT: the approximated loss regression with QAT could be turned into a classification problem. 
Rather than approximating the floating-point decision, one could try to obtain a yes/no answer to a different question: would the floating-point algorithm return an AD score larger than a threshold for this event?
In this way, one could set the threshold on the accurate floating-point model and could obtain good accuracy (in terms of anomaly acceptance) without having to predict the exact AD score value across multiple orders of magnitude.
\end{itemize}
In this paper, we focus on the first two approaches, leaving the investigation of the other approaches to future work.

This paper is structured as follows: in Section~\ref{sec:data} we describe the benchmark dataset. 
In Section~\ref{sec:AEmodels} a detailed description of the autoencoder models is given, followed by Section~\ref{sec:ADscores} in which the definition of the quantities used as anomaly detection scores is presented. 
In Section~\ref{sec:baseline} results of the uncompressed and unquantized model are presented. 
In the next part, Section~\ref{sec:compression}, model compression is detailed. 
Section~\ref{sec:fpga} describes the strategy to compress the models and deploy them on FPGAs, including an assessment of the required FPGA resources. 

\section{Data samples}
\label{sec:data}

This study follows the setup of Ref.~\cite{Cerri:2018anq,Nguyen:2018ugw}. 
We use a data sample that represents a typical proton-proton collision dataset that has been pre-filtered by requiring the presence of an electron or a muon with a transverse momentum $\pt>23$GeV and a pseudo-rapidity $|\eta|<3$~(electron) and $|\eta|<2.1$~(muon). This is representative of a typical L1T selection algorithm of a multipurpose LHC experiment. 
In addition to this, we consider the four benchmark new physics scenarios 
discussed in Ref.~\cite{Cerri:2018anq}:
\begin{itemize}
    \item A leptoquark (LQ) with a mass of 80\GeV, decaying to a $b$ quark and a $\tau$ lepton~\cite{zenodo_LQ},
    \item A neutral scalar boson ($A$) with a mass of 50\GeV, decaying to two off-shell Z
    bosons, each forced to decay to two leptons: $A \to 4\ell$~\cite{zenodo_AZZ},
    \item A scalar boson with a mass of 60\GeV, decaying to two tau leptons: $h^0\to \tau \tau$~\cite{zenodo_htautau},
    \item  A charged scalar boson with a mass of 60\GeV, decaying to a tau lepton and a neutrino: $h^\pm \to \tau \nu$~\cite{zenodo_htaunu}.
\end{itemize}
These four processes are used to evaluate the accuracy of the trained models.
A detailed description of the dataset can be found in Ref.~\cite{Govorkova:2021hqu}.

In total, the background sample consists of 8 million events. 
Of these, 50\% are used for training, 40\% for testing and 10\% for validation). 
The new physics benchmark samples are only used for evaluating the performance of the models.

The training dataset together with signal datasets for testing are published on Zenodo~\cite{zenodo_bkg_training, zenodo_LQ, zenodo_AZZ, zenodo_htautau, zenodo_htaunu}.

\section{Autoencoder models}
\label{sec:AEmodels}

We consider two classes of architectures: one based on dense feed-forward neural networks (DNNs) and one using convolutional neural networks (CNNs). 
Both start from the same inputs, namely the ($\pt$, $\eta$, $\phi$) 
values for 18 reconstructed objects (ordered as 4 muons, 4 electrons, and 10 jets), and the $\phi$ and magnitude of the missing transverse energy (MET), forming together an input of shape (19, 3) where MET $\eta$ values are zero-padded by construction ($\eta$ is zero for transverse quantities). For events with fewer than the maximum number of muons, electrons, or jets, the input is also zero-padded, as commonly done in the L1T algorithm logic.

\begin{figure*}[h!tb]
    \centering
    \includegraphics[width=0.75\textwidth]{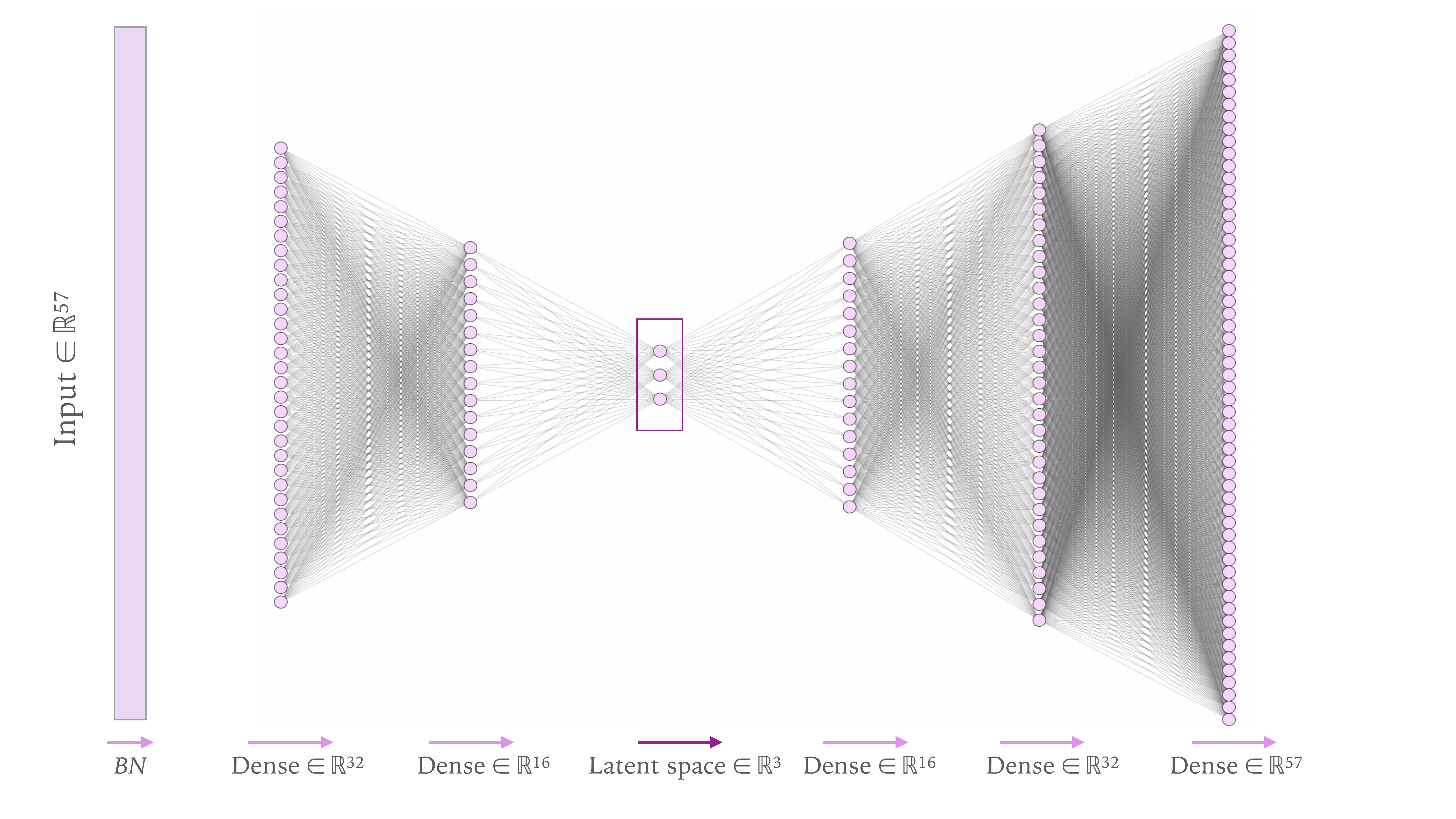}
    \includegraphics[width=0.75\textwidth]{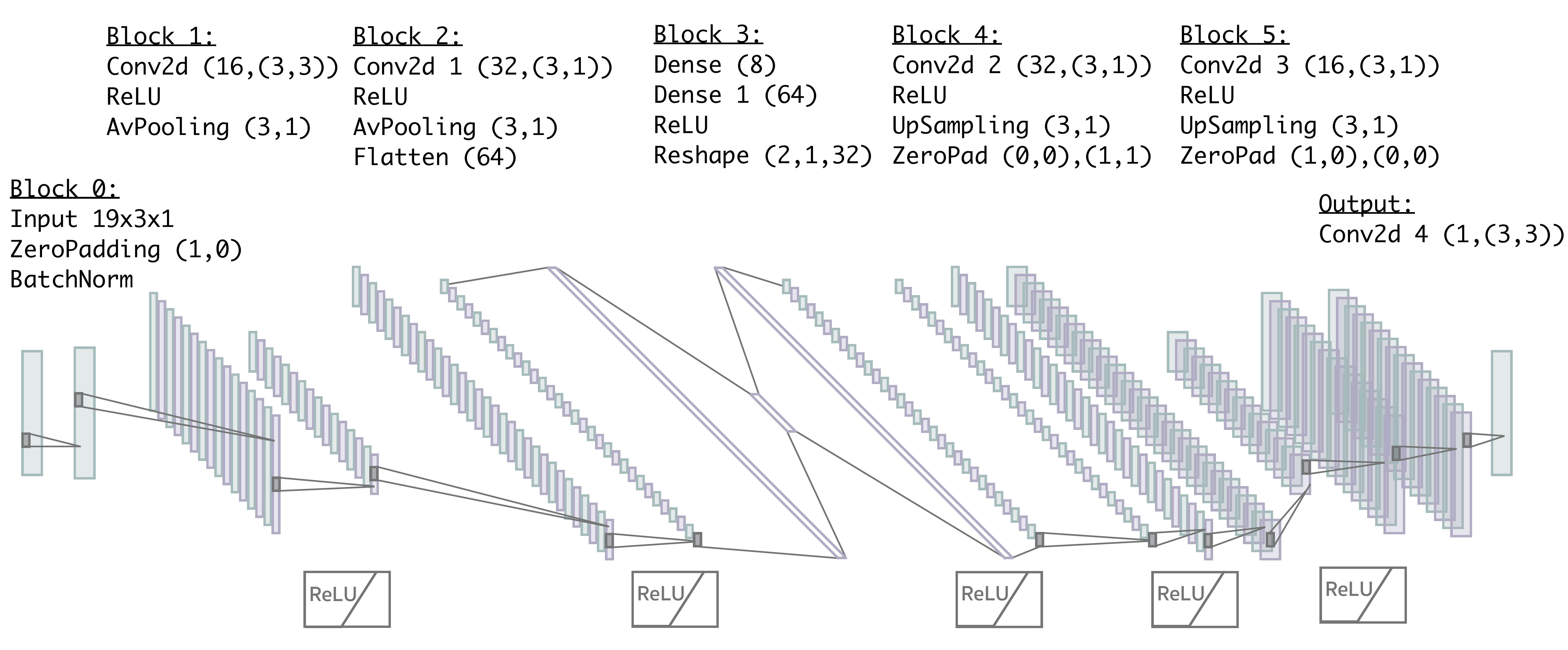}
    \caption{Network architecture for the DNN AE (top) and CNN AE (bottom) models. 
    The corresponding VAE models are derived introducing the Gaussian sampling in the latent space, for the same encoder and decoder architectures (see text).\label{fig:archs}}
\end{figure*}

In order to account for resource consumption and latency of the data pre-processing step, we use a batch normalization layer~\cite{batchnorm} as the first layer for each model.
As all processing is done on-chip, the resource and latency measurements will be consistent with those of a real L1T implementation.
For both architectures, CNN and DNN, we consider both a plain AE and a VAE. 
In the AE, the encoder provides directly the coordinates of the given input, projected in the latent space. 
In the VAE, the encoder returns the mean values $\vec{\mu}$ and the standard deviation $\vec{\sigma}$ of the $N$-dimensional Gaussian distribution which represents the latent-space probability density function associated with a given event.

For the DNN model, the four-vector of each reconstructed object is flattened and concatenated into a 1D array, resulting in a 57-dimension input vector. 
The DNN AE architecture is shown on the top plot in Figure~\ref{fig:archs}. 
All of the inputs are batch-normalized and passed through a stack of 3 fully connected layers, with 32, 16, and 3 nodes. 
The output of each layer is followed by a batch normalization layer and activated by a leaky ReLU function~\cite{leakyrelu}. 
The 3-dimensional output of the encoder is the projection of the input in the latent space. 
The decoder consists of a stack of 3 layers, with 16, 32, and 57 nodes. 
As for the encoder, we use a batch normalization layer between the fully connected layer and its activation. 
The last layer has no activation function, while leaky ReLU is used for the others. 
The DNN VAE follows the same architecture, except for the latent-space processing, which follows the usual VAE prescription: two 3-dimensional fully connected layers produce the $\vec{\mu}$ and $\vec{\sigma}$ vectors from which Gaussian latent quantities are sampled and injected in the decoder.

The CNN AE architecture is shown on the bottom plot in Figure~\ref{fig:archs}. 
The encoder takes as input the single-channel 2D array of four-momenta including the two MET-related features (magnitude and $\phi$ angle) and zeros for MET $\eta$, resulting in a total input size of $19\times3\times1$. 
It should be emphasised that we are not using image data, rather treating tabular data as a 2D image to make it possible to explore CNN architectures.
The input is first zero-padded in order to resize the image to $20\times3\times1$, which is required in order to parallelize the network processing in the following layer on the FPGA.
For the Conv2D FPGA implementation, we control how many iterations of outer loop (over the rows of the image array) are running in parallel. 
To simplify the implementation we run the same number of iterations in parallel, which requires that the number of rows in the input image is an integer multiple of the number of parallel processors.
Since 19 is a prime number, we choose to extend the input size to 20 before passing it through the Conv2D layer.
After padding, the input is scaled by a batch normalization layer and then processed by a stack of two CNN blocks, each including a 2D convolutional layer followed by a ReLU~\cite{conf/icml/NairH10} activation function. 
The first layer has 16 $3\times3$ kernels, without padding to ensure that $\pt, \eta$ and $\phi$ inputs do not share weights. 
The second layer has 32 $3\times1$ kernels. 
Both layers have no bias parameters and a stride set to one. 
The output of the second CNN block is flattened and passed to a DNN layer, with 8 neurons and no activation, which represents the latent space. 
The decoder takes this as input to a dense layer with 64 nodes and ReLU activation, and reshapes it into a $2\times1\times32$ table. 
The following architecture mirrors the encoder architecture with 2 CNN blocks with the same number of filters as in the encoder and with ReLU activation. 
Both are followed by an upsampling layer, in order to mimic the result of a transposed convolutional layer. 

Finally, one convolutional layer with a single filter and no activation function is added.
Its output is interpreted as the AE reconstructed input. 
The CNN VAE is derived from the AE, including the $\vec{\mu}$ and $\vec{\sigma}$ Gaussian sampling in the latent space.

All models are implemented in \Tensorflow, and trained on the background dataset by minimizing 
a customized mean squared error (MSE) loss with the Adam~\cite{adam} optimizer.
In order to aid the network learning process, we use a dataset with standardized $\pt$ as a target, so that all the quantities are $\mathcal{O}(1)$. 
To account for physical boundaries of $\eta$ and $\phi$, for those features a re-scaled \texttt{tanh} activation is used in the loss computation.
In addition, the sum in the MSE loss is modified in order to ignore the zero-padding entries of the input dataset and the corresponding outputs. 
When training the VAE, the loss is changed to:
\begin{equation}
    {\cal L} = (1-\beta) \mathrm{MSE}(\mathrm{Output}, \mathrm{Input}) + \beta \mathrm{D_\mathrm{{KL}}}(\vec{\mu}, \vec{\sigma})~,
    \label{eq:vae_loss}
\end{equation}
where MSE labels the reconstruction loss (also used in the AE training), 
\Dkl is the Kullback-Leibler regularization term~\cite{Joyce2011} usually adopted for VAEs
\begin{equation}
    \mathrm{D_\mathrm{{KL}}}(\vec{\mu}, \vec{\sigma}) = -\frac{1}{2}\sum_i \left ( \log(\sigma_i^2) - \sigma_i^2 -\mu_i^2 +1 \right)~,
    \label{eq:D_KL}
\end{equation}
and $\beta$ is a hyperparameter defined in the range $[0, 1]$~\cite{betaVAE}.

Both models are trained for 100 epochs with a batch size of 1024, using early stopping if there 
is no improvement in the loss observed after ten epochs. 
All models are trained with floating point precision on an NVIDIA RTX2080 GPU. 
We refer to these as the baseline floating-point (BF) models. 

\section{Anomaly detection scores}
\label{sec:ADscores}
An autoencoder is optimized to retain the minimal set of information needed to reconstruct a accurate estimate of the input. 
During inference, an autoencoder might have problems generalizing to features it was not exposed to during training. 
Selecting events where the autoencoder output is far from the given input is often seen as an effective AD algorithm. 
For this purpose, one could use a metric that measures the distance between the input and the output. 
The simplest solution is to use the same metric that defines the training loss function.
In our case, we use the MSE between the input and the output. 
We refer to this strategy as \textit{input-output} (IO) AD.

In the case of a VAE deployed in the L1T, one cannot simply exploit an IO AD strategy since this would require sampling random numbers on the FPGA. 
The trigger decision would not be deterministic, something usually tolerated only for service triggers, and not for triggers serving physics studies. 
Moreover, one would have to store random numbers on the FPGA, which would consume resources and increase the latency. 
To deal with this problem, we consider an alternative strategy by defining an AD score based on the $\vec{\mu}$ and $\vec{\sigma}$ values returned by the encoder (see Eq.~(\ref{eq:vae_loss})).
In particular, we consider two options: the KL divergence term entering the VAE loss (see Eq.~(\ref{eq:D_KL})) and the z-score of the origin $\vec{0}$ in the latent space with respect to a Gaussian distribution centered at $\vec{\mu}$ with standard deviation $\vec{\sigma}$~\cite{aarrestad2021dark}:
\begin{equation}
    R_z = \sum_i \frac{\mu_i^2}{\sigma_i^2}.
\end{equation}
These two AD scores have several benefits we take advantage of: Gaussian sampling is avoided; we save significant resources and latency by not evaluating the decoder; and we do not need to buffer the input data for computation of the MSE. 
During the model optimization, we tune $\beta$ so that we obtain (on the benchmark signal models) comparable performance for the \Dkl AD score and the IO AD score of the VAE.

\section{Performance at floating-point precision}
\label{sec:baseline}

\begin{figure*}[h!tb]
    \centering
    \includegraphics[width=0.4\textwidth,trim=15 25 15 10]{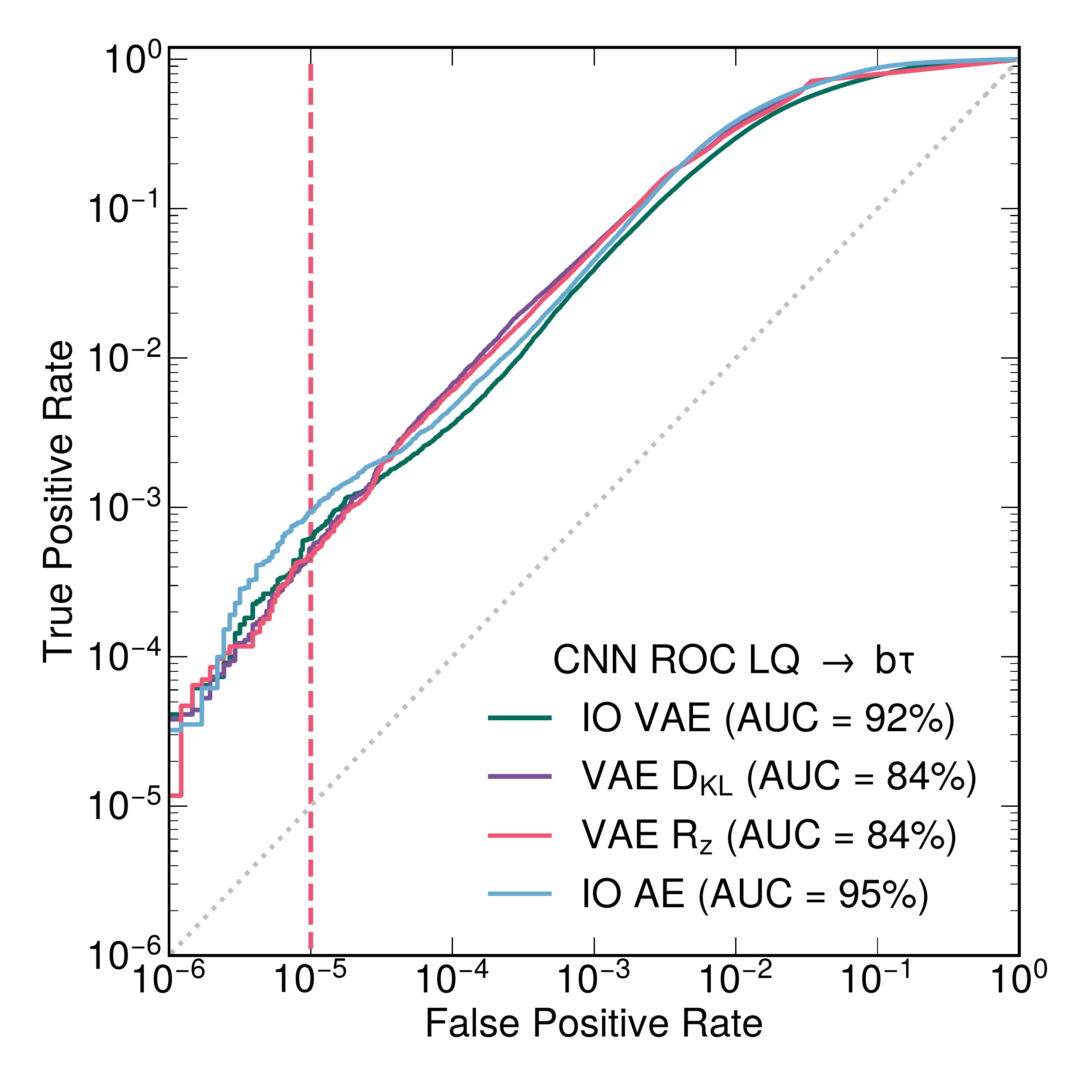}
    \includegraphics[width=0.4\textwidth,trim=15 25 15 10]{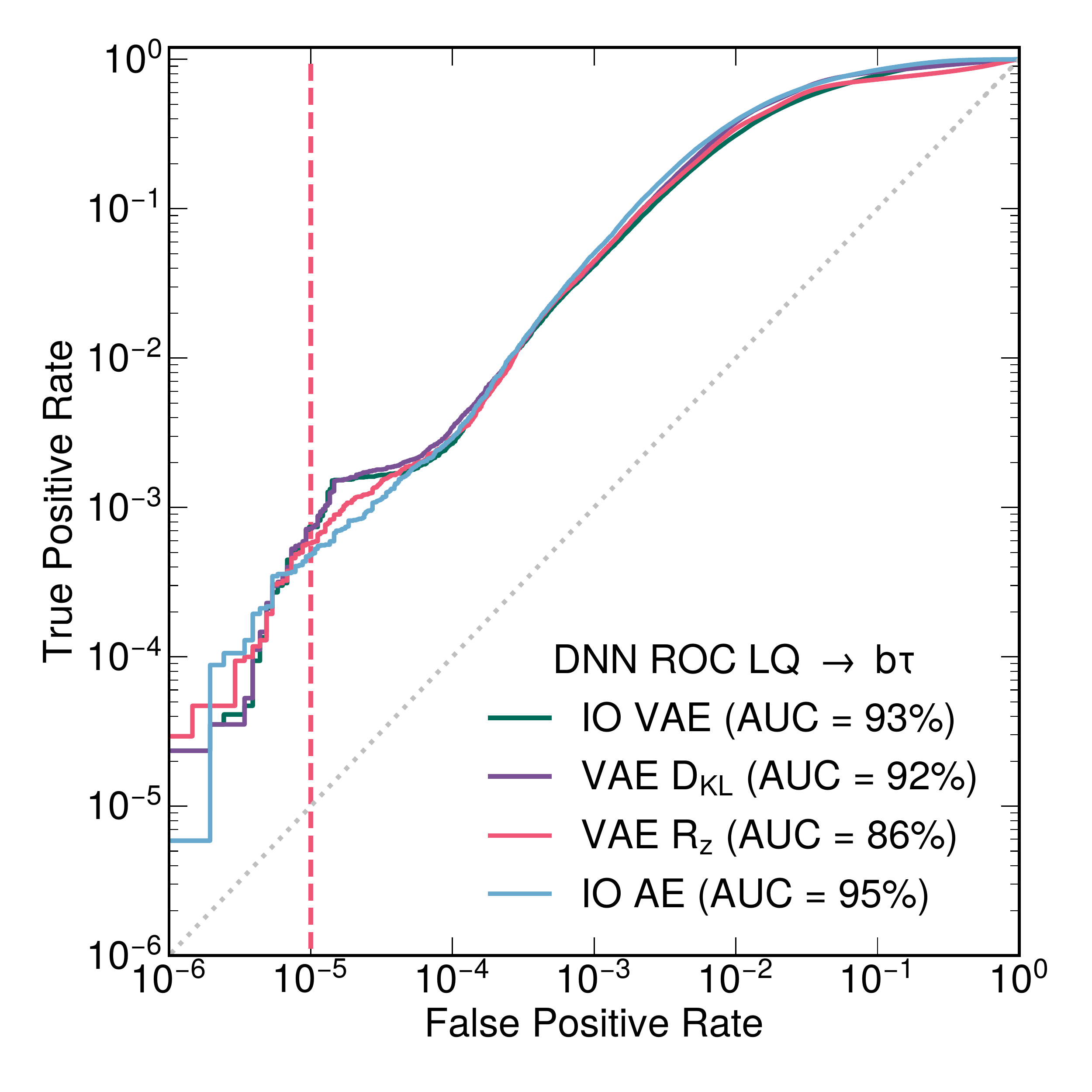}
    \includegraphics[width=0.4\textwidth,trim=15 25 15 10]{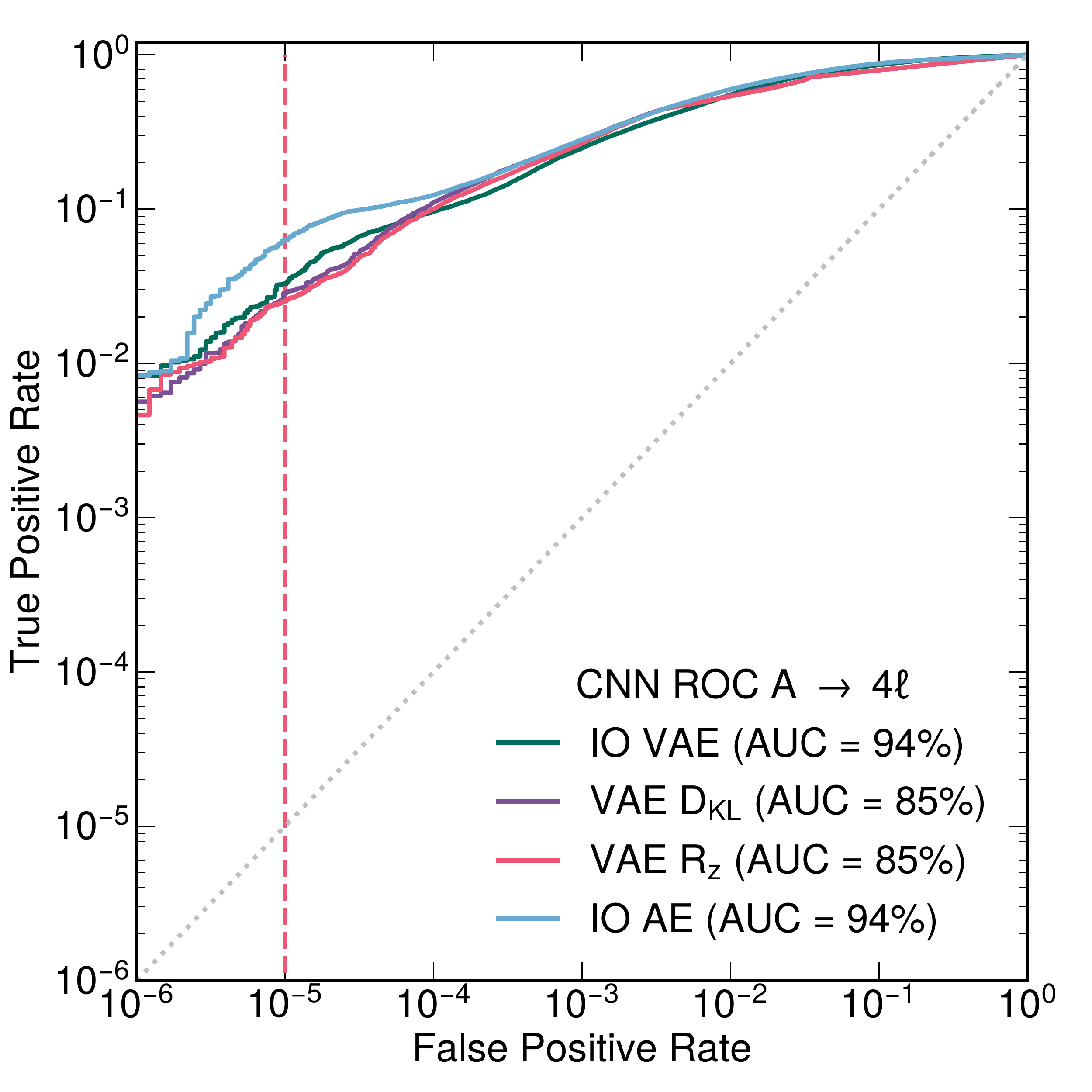}
    \includegraphics[width=0.4\textwidth,trim=15 25 15 10]{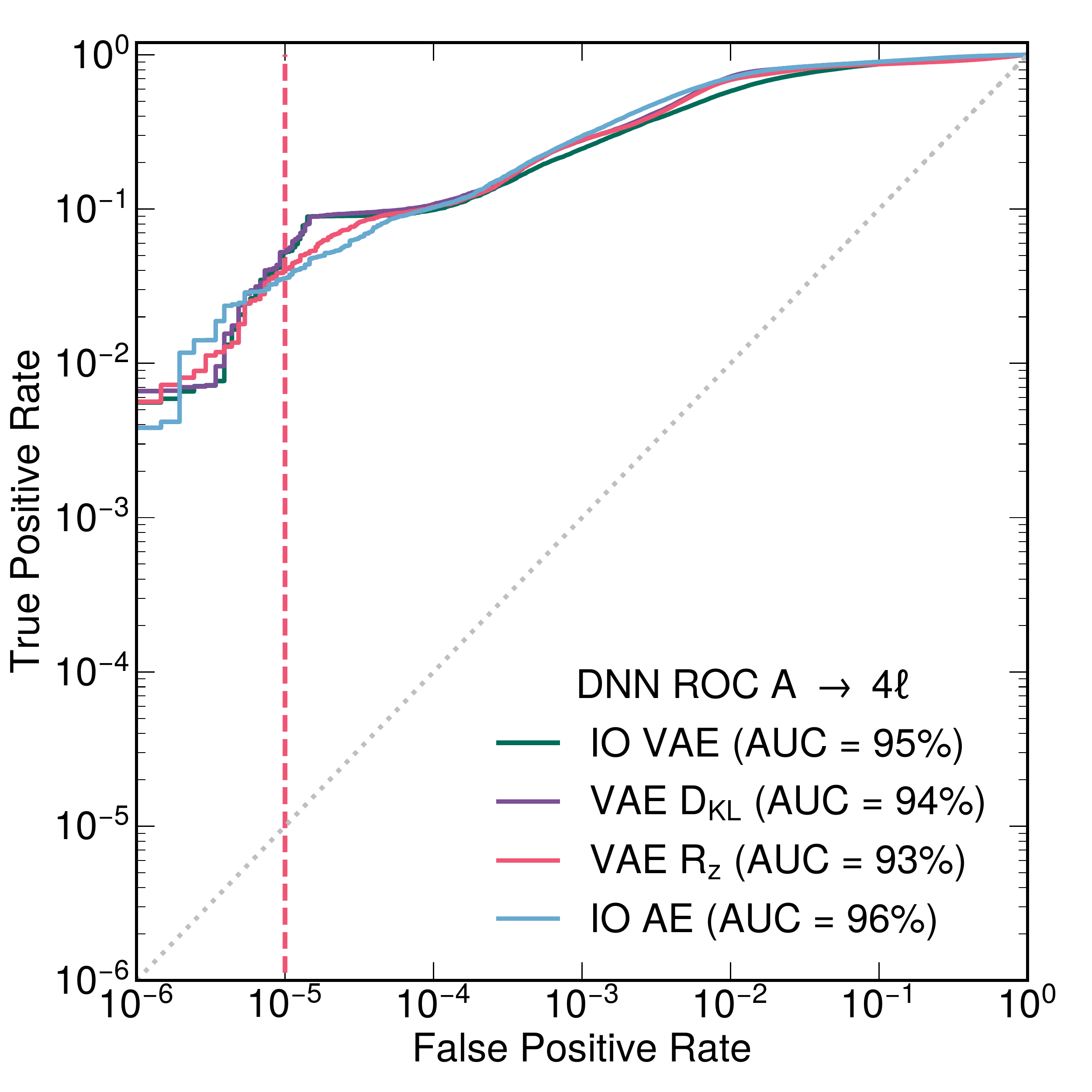}
    \caption{ROC curves of four AD scores (IO AD for AE and VAE models, \Rz and \Dkl ADs for the VAE models) for the CNN (left) and DNN (right) models, obtained from the two new physics benchmark models: $LQ\to b\tau$ (top) and $A\to4\ell$ (bottom).\label{fig:ROC}}
\end{figure*}

\begin{figure*}[h!tb]
    \centering
        \includegraphics[width=0.4\textwidth,trim=10 10 10 10]{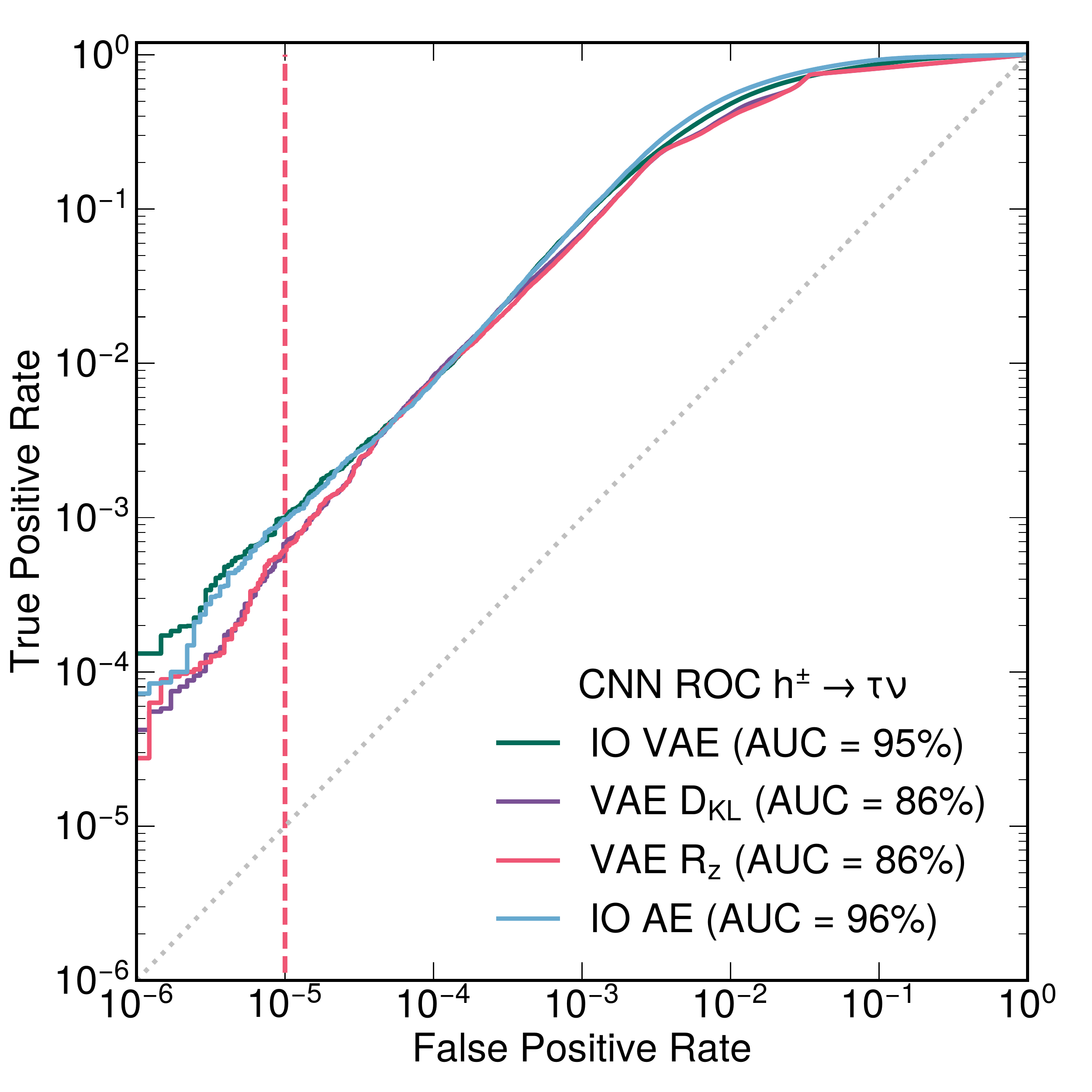}
        \includegraphics[width=0.4\textwidth,trim=10 10 10 10]{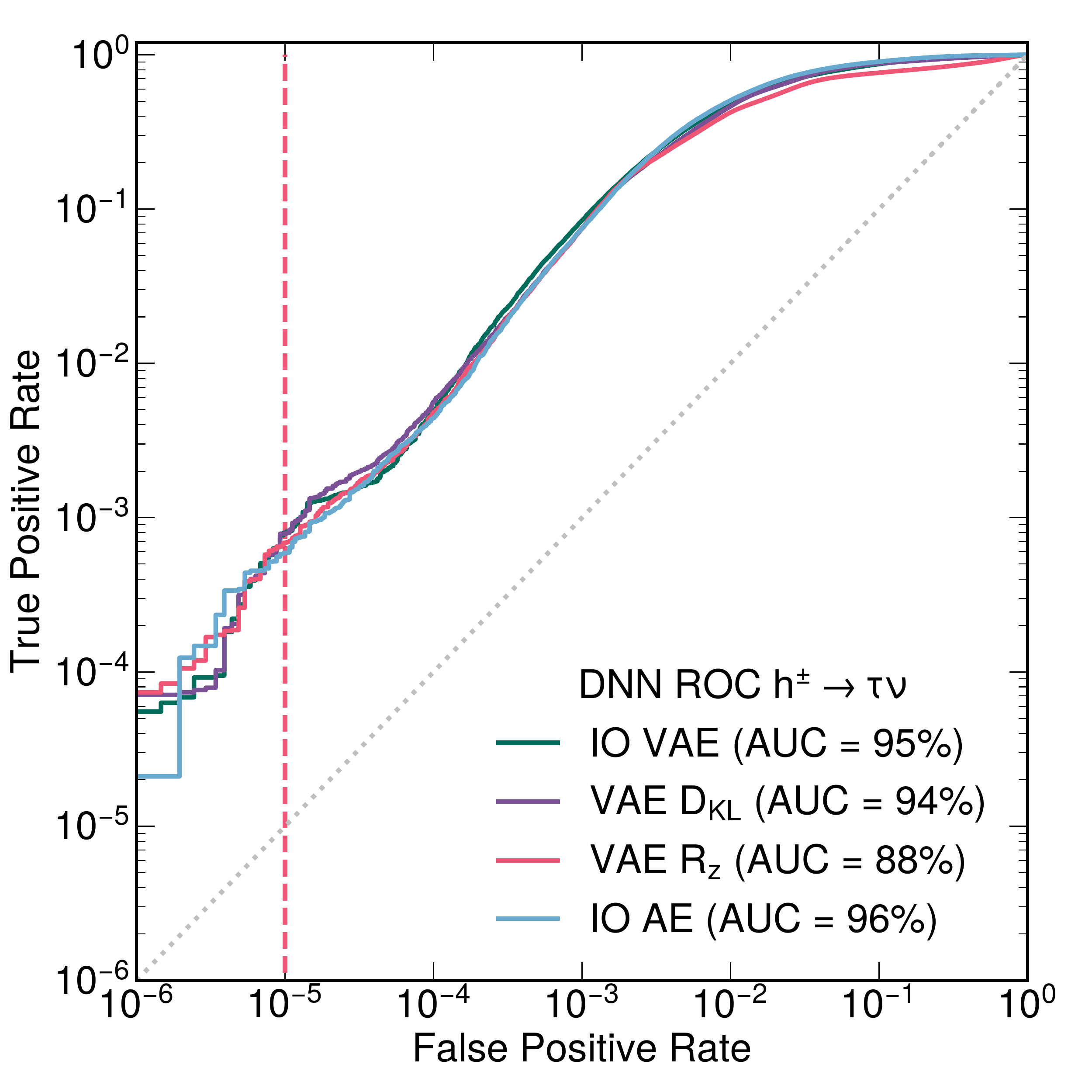}
        \includegraphics[width=0.4\textwidth,trim=10 10 10 10]{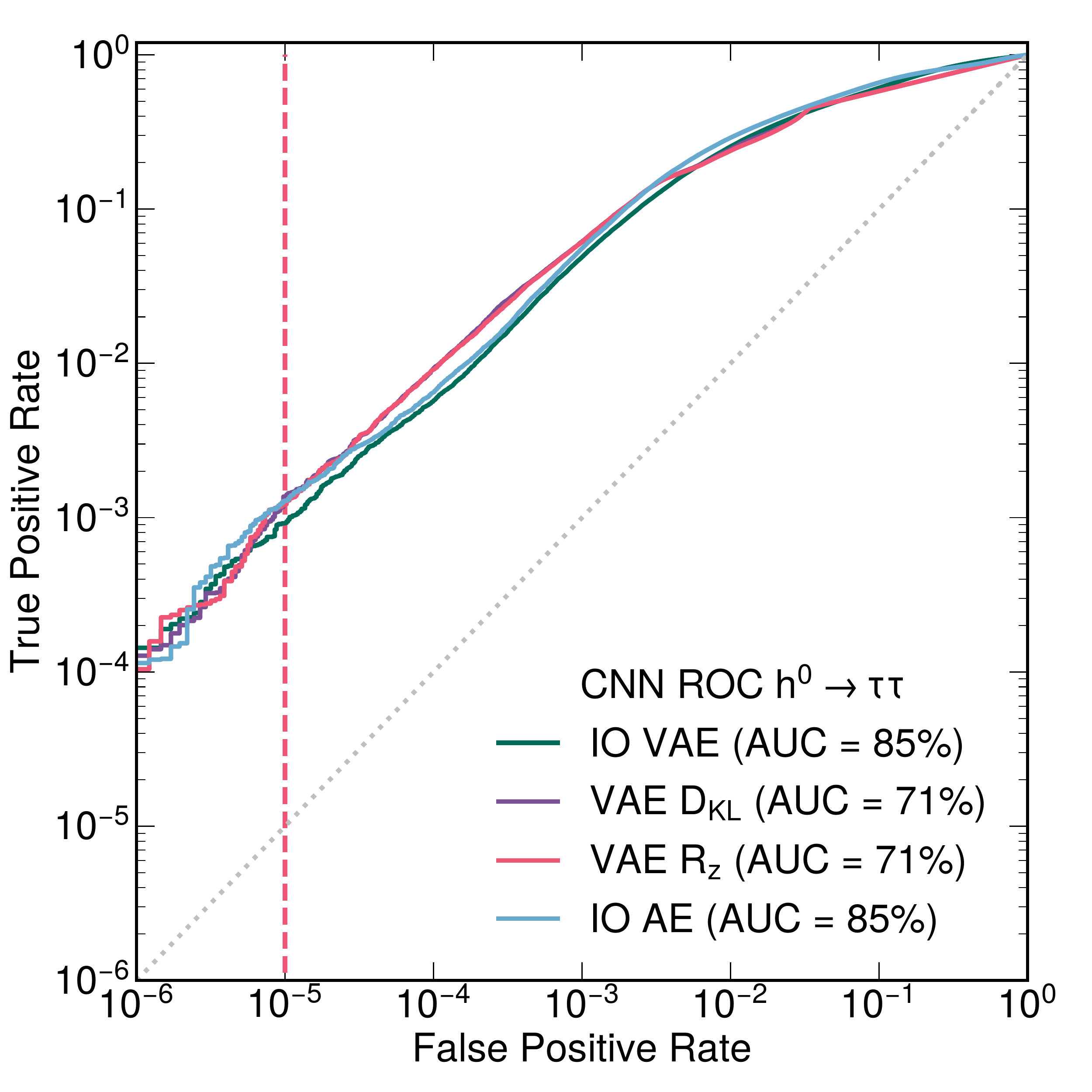}
        \includegraphics[width=0.4\textwidth,trim=10 10 10 10]{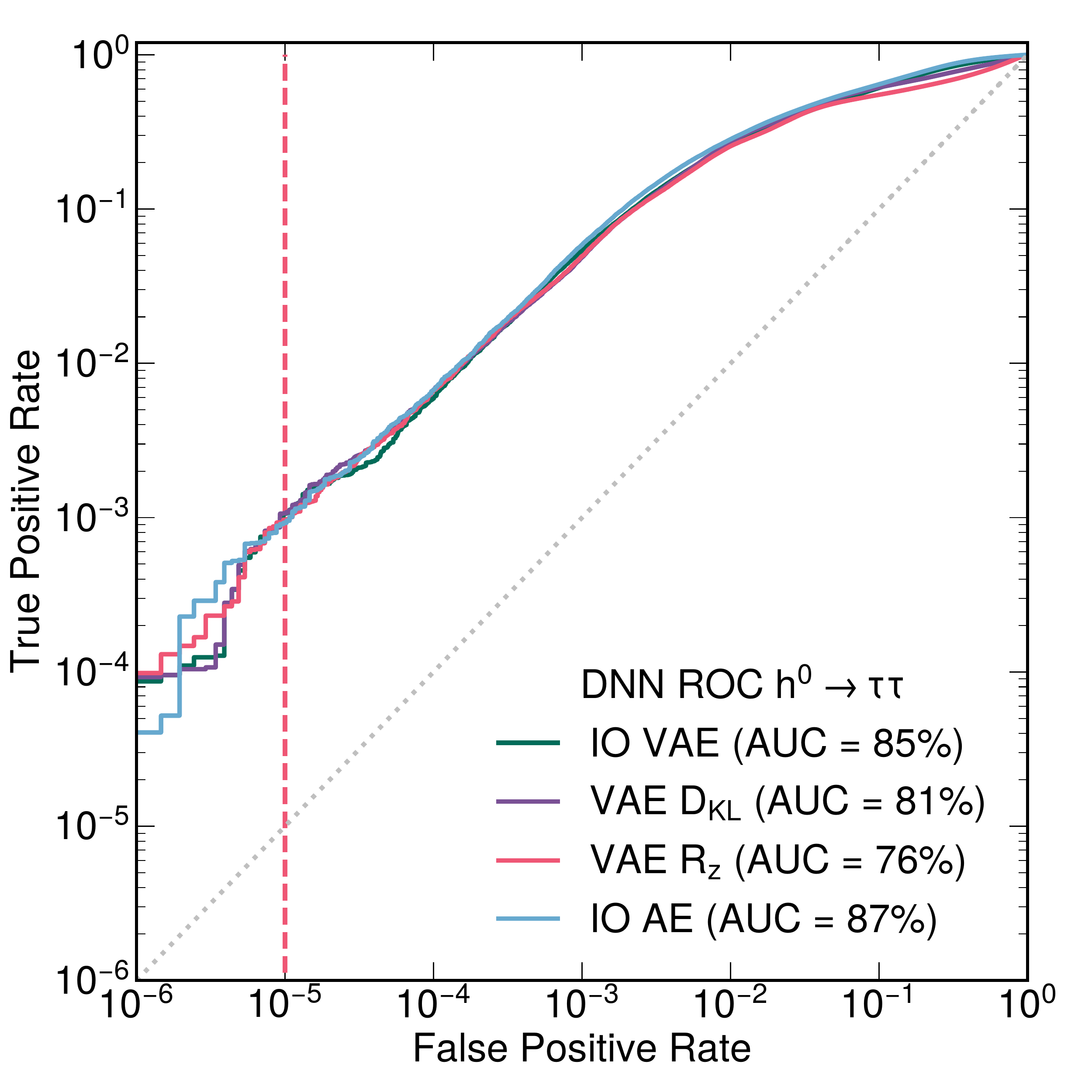}
    \caption{ROC curves of four AD scores (IO AD for AE and VAE models, \Rz and \Dkl ADs for the VAE models) for the CNN (left) and DNN (right) models, obtained from two new physics benchmark models: $h^{\pm}\to\tau\nu$ (top) and $h^{0}\to\tau\tau$ (bottom). \label{fig:ROC2}}
\end{figure*}

The model performance is assessed using the four new physics benchmark models. 
The anomaly-detection scores considered in this paper are IO AD for the AE models, \Rz and \Dkl ADs for the VAE models.
For completeness, results obtained from the IO AD score of the VAE models are also shown. 
The receiver operating characteristic (ROC) curves in Figures~\ref{fig:ROC} and \ref{fig:ROC2} 
show the true positive rate (TPR) as a function of the false positive rate (FPR), computed by changing the lower threshold applied on the different anomaly scores.
We further quantify the AD performance quoting the area under the ROC curve (AUC) and the TPR corresponding to a FPR working point of $10^{-5}$~(see Table~\ref{tab:performance_baseline}), which on this dataset corresponds to the reduction of the background rate to approximately 1000 events per month.

From the ROC curves, we conclude that \Dkl can be used as an anomaly metric for both the DNN and CNN VAE. This has the potential to significantly reduce the inference latency and on-chip resource consumption as only half of the network (the encoder) needs to be evaluated and that there no longer is a need to buffer the input in order to compute an MSE loss. The \Rz metric performs worse and is therefore not included in the following studies. 

\begin{table*}[tb]
    \centering
    \caption{Performance assessment of the CNN and DNN models, for different AD scores and different new physics benchmark scenarios.\label{tab:performance_baseline}}
    \begin{tabular}{l|c|c|c|c|c|c|c|c|c}
    \hline
    \multirow{2}{*}{Model}  & \multirow{2}{*}{AD score}  & \multicolumn{4}{|c|}{TPR @ FPR $10^{-5}$[\%]} & \multicolumn{4}{|c}{AUC[\%]} \\
    \cline{3-10}
     & & LQ $\rightarrow$ b$\tau$ &$A \rightarrow 4\ell$ & $h^{\pm} \rightarrow \tau\nu$  & $h^{0} \rightarrow \tau\tau$ & LQ $\rightarrow$ b$\tau$ &$A \rightarrow 4\ell$& $h^{\pm} \rightarrow \tau\nu$  & $h^{0} \rightarrow \tau\tau$ \\
\hline
\hline
\multirow{3}{*}{CNN VAE} & IO & 0.06&3.28&0.10&0.09& 92&94&95&85\\
\cline{2-10}
CNN VAE & \Dkl & 0.05&2.85&0.07&0.14& 84&85&86&71\\
\cline{2-10}
        & \Rz  & 0.05&2.53&0.06&0.12& 84&85&86&71\\
\hline
CNN AE  & IO & 0.09&6.29&0.10&0.13& 95&94&96&85\\
\hline
\hline
\multirow{3}{*}{DNN VAE} & IO & 0.07&5.23&0.08&0.11& 93&95&95&85\\
\cline{2-10}
DNN VAE & \Dkl & 0.07&5.27&0.08&0.11& 92&94&94&81\\
\cline{2-10}
        & \Rz  & 0.06&4.05&0.07&0.10& 86&93&88&76\\
\hline
DNN AE  & IO & 0.05&3.56&0.06&0.09& 95&96&96&87\\
\hline
\hline
    \end{tabular}
\end{table*}

\section{Model compression}
\label{sec:compression}
We adopt different strategies for model compression.
First of all, we compress the BF model by pruning the dense and convolutional layers by 50\% of their connections, following the same procedure as Ref.~\cite{aarrestad2021fast}. 
Pruning is enforced using the polynomial decay implemented in \Tensorflow pruning API, a \Keras-based~\cite{keras} interface consisting of a simple drop-in replacement of \Keras layers. 
A sparsity of 50\% is targeted, meaning only 50\% of the weights are retained in the pruned layers and the remaining ones are set to zero. 
The pruning is set to start from the fifth epoch of the training to ensure the model is closer to a stable minimum before removing weights deemed unimportant.
By pruning the BF model layers to a target sparsity of 50\%, the number of floating-point operations required when evaluating the model, can be significantly reduced. 
We refer to the resulting model as the baseline pruned (BP) model.
For the VAE, only the encoder is pruned, since only that will be deployed on FPGA. 
The BP models are taken as a reference to evaluate the resource saving of the following compression strategies, including QAT and PTQ.

Furthermore, we perform a QAT of each model described in Section~\ref{sec:AEmodels}, implementing them in the \QKeras library~\cite{AutoQ}. 
The bit precision is scanned between 2 and 16 with a 2-bit step. 
When quantizing a model, we also impose a pruning of the dense (convolutional) 
layers by 50\%, as done for the DNN (CNN) BP models. 
The results of QAT are compared to results obtained by applying a fixed-point precision to a BP floating-point model (i.e. using PTQ), using the same bit precision scan. 

\begin{figure*}[tb]
    \centering
        \includegraphics[width=0.4\textwidth,trim=10 0 10 10]{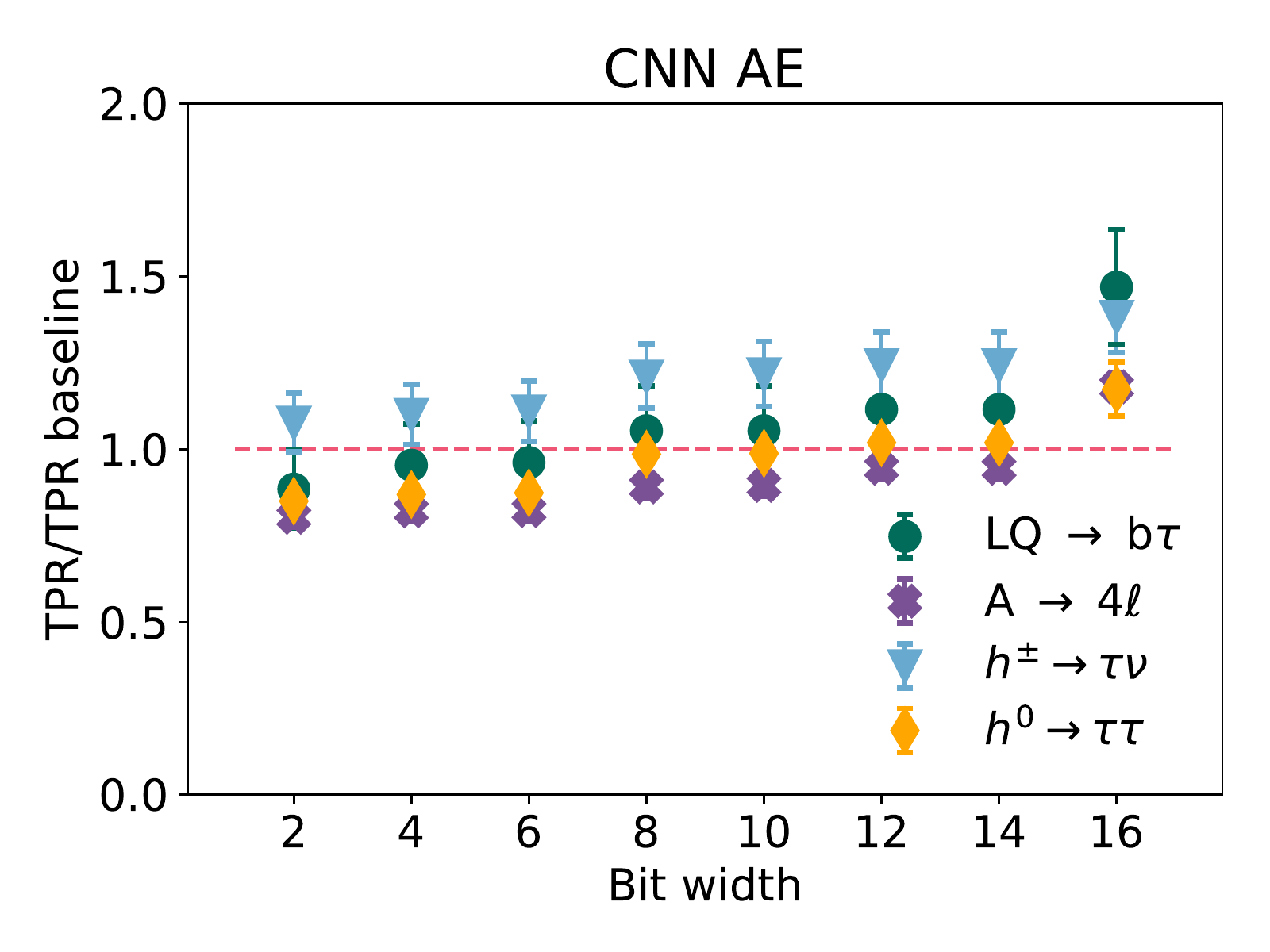} 
        \includegraphics[width=0.4\textwidth,trim=10 0 10 10]{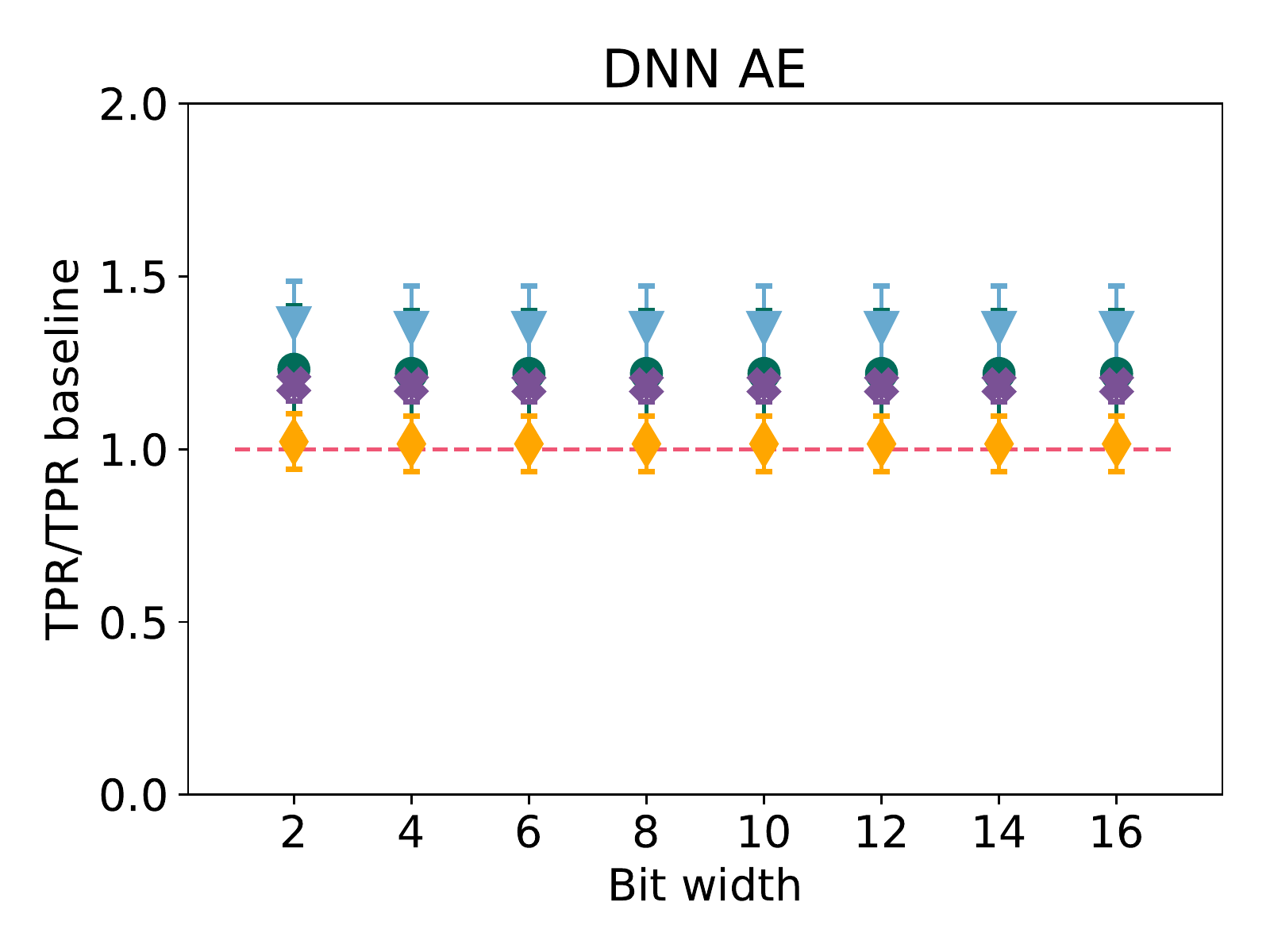} 
        \includegraphics[width=0.4\textwidth,trim=10 0 10 10]{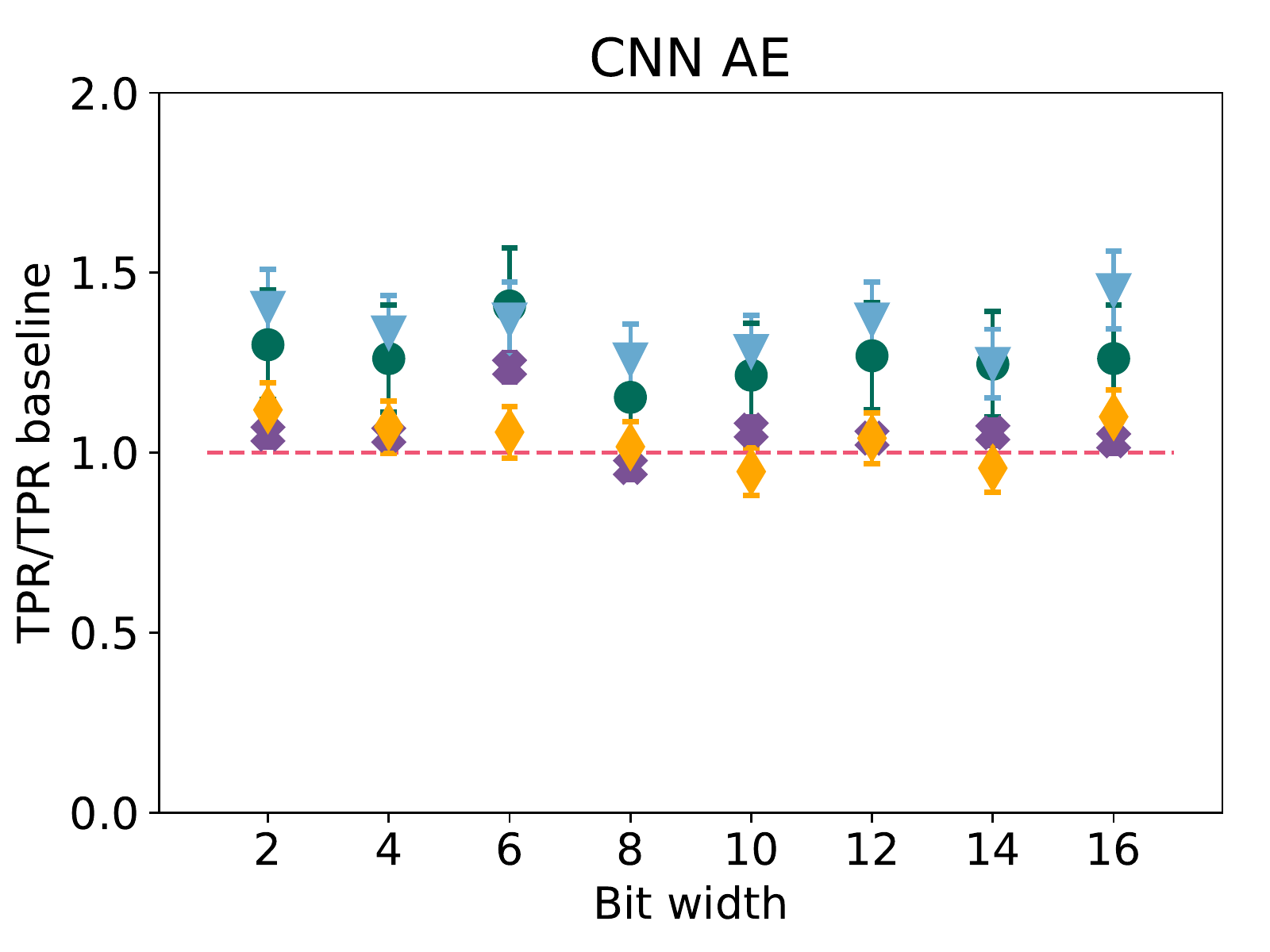}  \includegraphics[width=0.4\textwidth,trim=10 0 10 10]{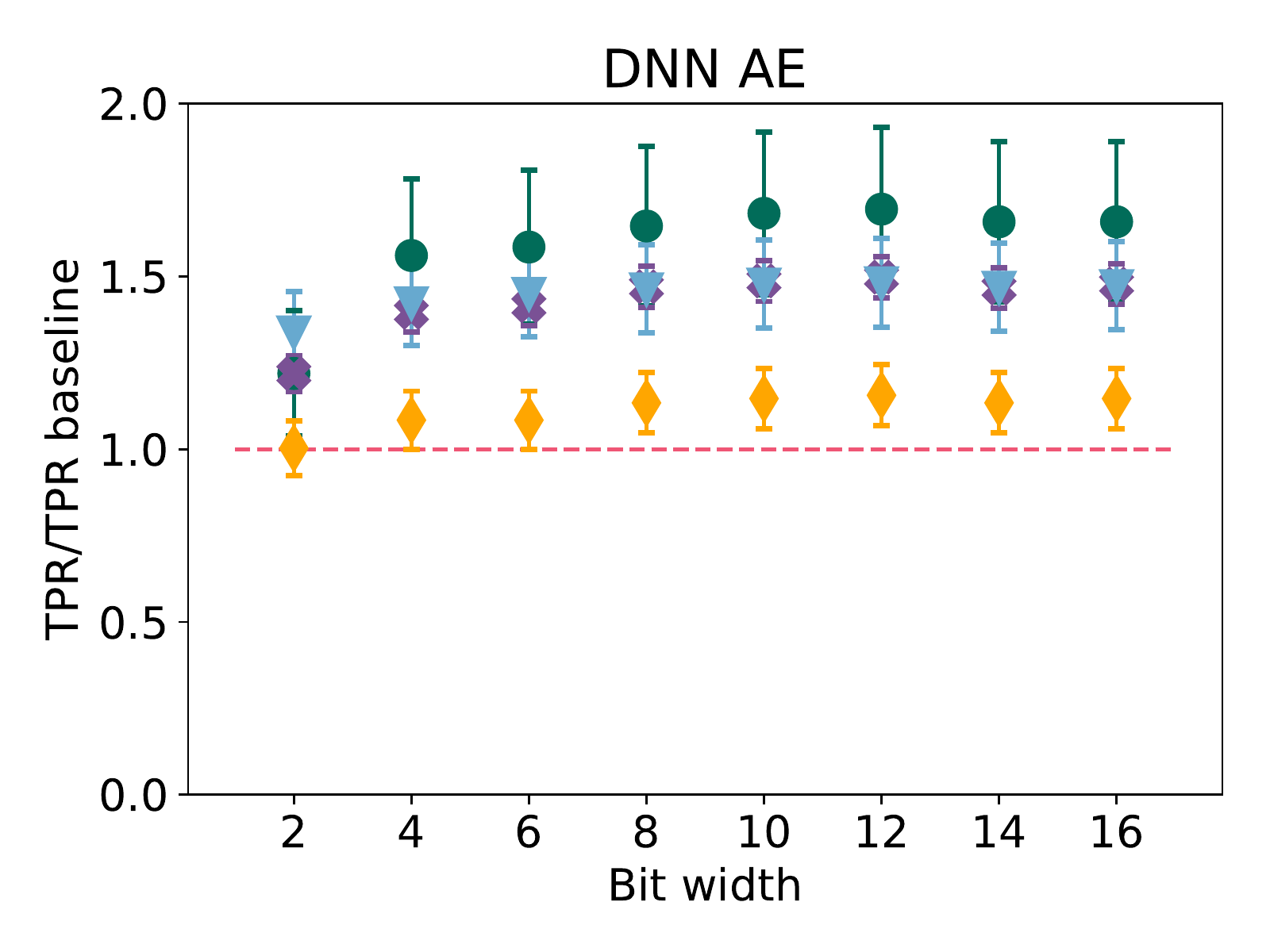} 
    \caption{TPR ratios versus model bit width for the AE CNN (left) and DNN (right) models tested on four new physics benchmark models, using mean squared error as figure of merit for PTQ (top) and QAT (bottom) strategies.}
    \label{fig:ratio_quant_ae}
\end{figure*}

\begin{figure*}[tb]
  \centering
        \includegraphics[width=0.4\textwidth,trim=10 0 10 10]{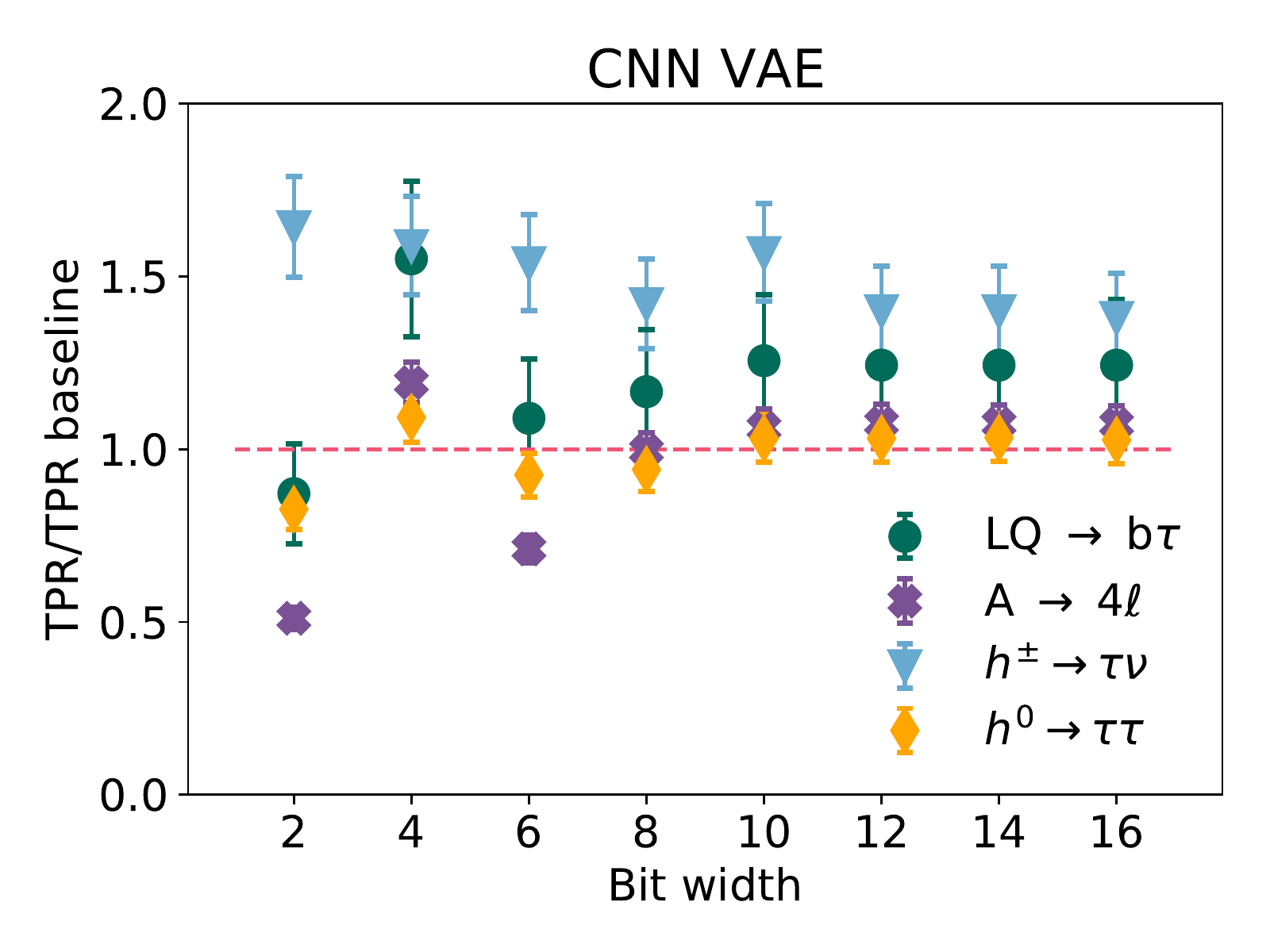}  \includegraphics[width=0.4\textwidth,trim=10 0 10 10]{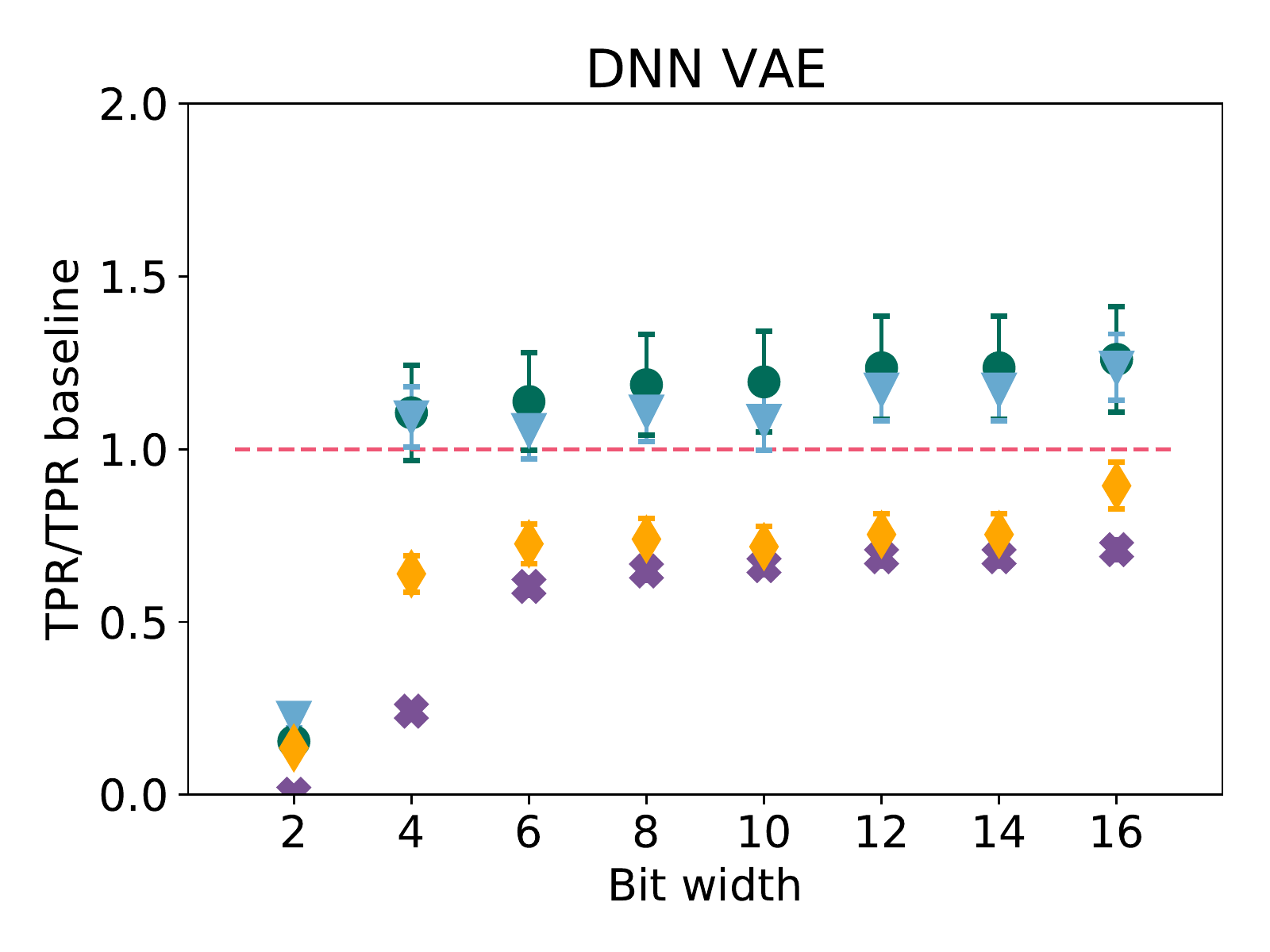}
        \includegraphics[width=0.4\textwidth,trim=10 0 10 10]{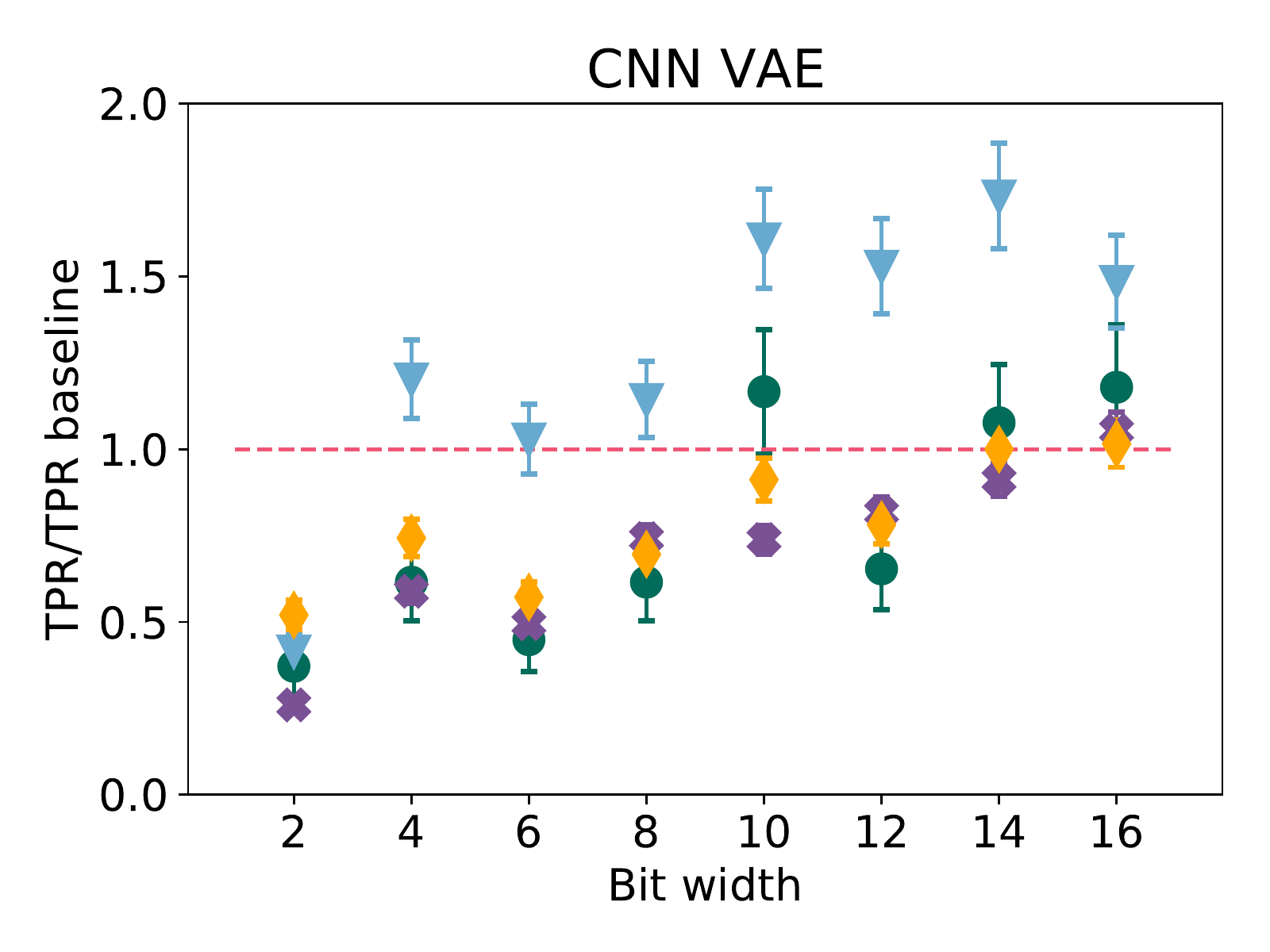}
        \includegraphics[width=0.4\textwidth,trim=10 0 10 10]{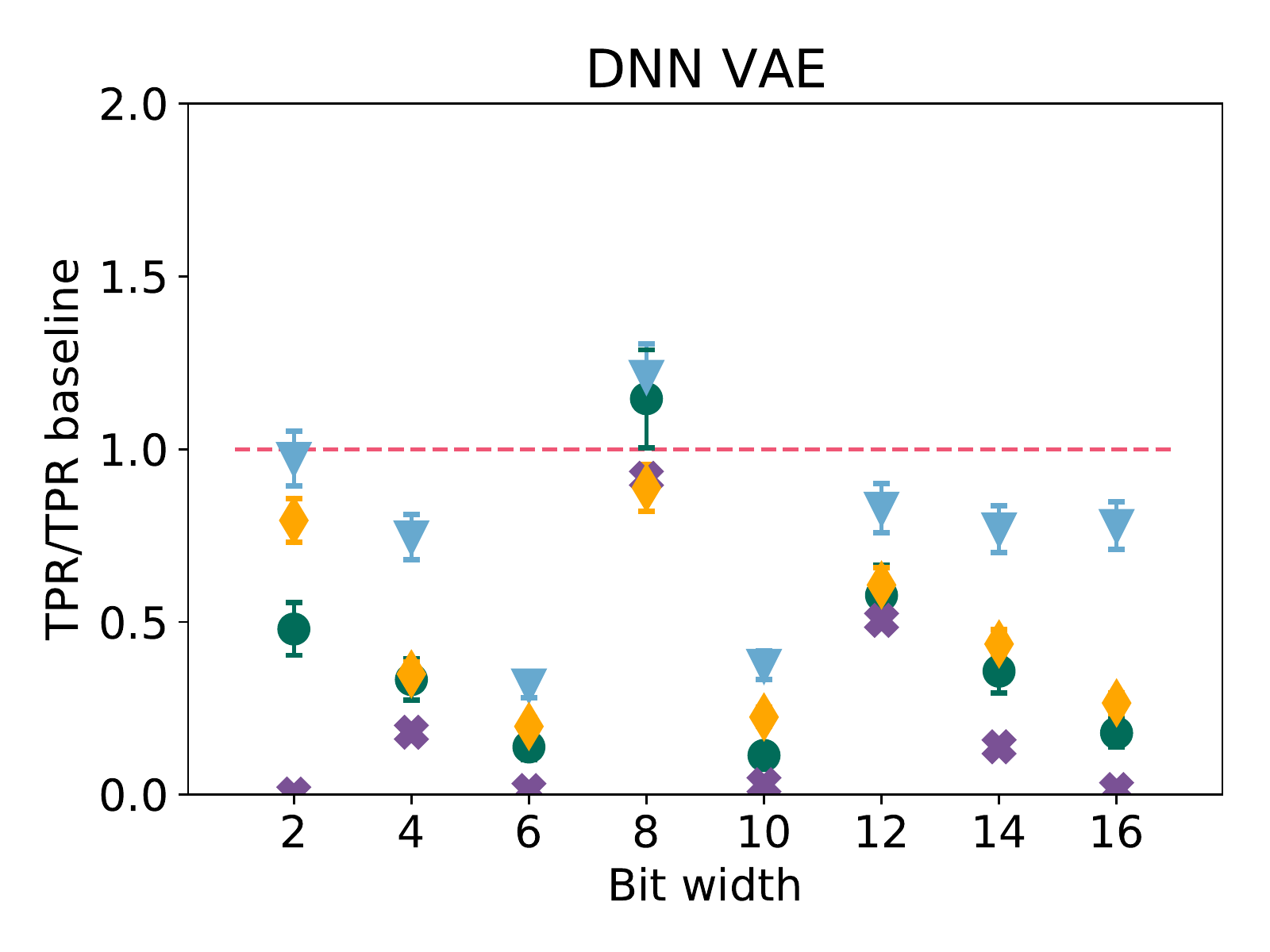}
    \caption{TPR ratios versus model bit width for the VAE CNN (left) and DNN (right) models tested on four new physics benchmark models, using \Dkl~as figure of merit for PTQ (top) and QAT (bottom) strategies.}
    \label{fig:ratio_quant_vae_kl}
\end{figure*}

Performance of the quantized models, both for QAT and PTQ, is assessed using the TPR obtained for an FPR of $10^{-5}$ for the given precision. 
The bottom plots in Figures~\ref{fig:ratio_quant_ae} and \ref{fig:ratio_quant_vae_kl} show ratios of QAT performance quantities obtained for each bit width with respect to the BP model performance of the AE and VAE, respectively. 
The top plots show ratios of PTQ performance quantities obtained in the same manner as for QAT.

Based on these ratio plots, the precision used for the final model is chosen. 
As expected, the performance of the VAEs is not stable as a function of bit width, since the AD figure of merit used for inference (\Dkl) is different from those minimized during the QAT training (VAE IO).
Therefore, we use PTQ compression for both DNN and CNN VAEs because they show stable results as a function of the bit width. 
For DNN and CNN VAE both PTQ and QAT show stable results, and therefore we choose QAT for AEs.
For the QAT CNN, the QAT DNN AE and the PTQ DNN VAE a bit width of 8 is chosen, and for the PTQ CNN VAE a bit width of 4 is used. 
The performance numbers for the chosen models are summarized in Table~\ref{tab:performance_quantised}.

\begin{table*}[tb]
    \centering
    \caption{Performance assessment of the quantized and pruned CNN and DNN models, for different AD scores and different new physics benchmark scenarios.\label{tab:performance_quantised}}
    \begin{tabular}{l|c|c|c|c|c|c|c|c|c}
    \hline
    \multirow{2}{*}{Model}  & \multirow{2}{*}{AD score}  & \multicolumn{4}{|c|}{TPR @ FPR $10^{-5}$[\%]} & \multicolumn{4}{|c}{AUC[\%]} \\
    \cline{3-10}
     & & LQ $\rightarrow$ b$\tau$ &$A \rightarrow 4\ell$ & $h^{\pm} \rightarrow \tau\nu$  & $h^{0} \rightarrow \tau\tau$ & LQ $\rightarrow$ b$\tau$ &$A \rightarrow 4\ell$ & $h^{\pm} \rightarrow \tau\nu$  & $h^{0} \rightarrow \tau\tau$ \\
\hline
\hline
CNN AE QAT 4 bits & IO & 0.09&5.96&0.10&0.13& 94&96&96&88\\
\hline
CNN VAE PTQ 8 bits  & \Dkl & 0.05&2.56&0.06&0.12& 84&84&85&71\\
\hline
\hline
DNN AE QAT 8 bits & IO & 0.08&5.48&0.09&0.11& 95&96&96&88\\
\hline
DNN VAE PTQ 8 bits & \Dkl & 0.08&3.41&0.09&0.08& 92&94&94&81\\
\hline
\hline
    \end{tabular}
\end{table*}

\section{Porting the algorithm to FPGAs}
\label{sec:fpga}
The models described above are translated into firmware using \hlsfml, then synthesized with Vivado HLS 2020.1~\cite{vivadohls}, targeting a Xilinx Virtex UltraScale+ VU9P (\texttt{xcvu9p-flgb2104-2-e}) FPGA with a clock frequency of 200\unit{MHz}. 
In order to have fair resource and latency estimations, obtained from the HLS C Simulation we have implemented custom layers in \hlsfml, which in the case of AE computes the loss function between the input and network output and for VAE computes the \Dkl term of the loss. 

A summary of the accuracy, resource consumption, and latency for the QAT DNN and CNN BP AE models, 
and the PTQ DNN and CNN BP VAE models is shown in Table~\ref{tab:resources}. 
Resource utilization is quoted as a fraction of the total available resources on the FPGA. 
We find the resources are less than about 12\% of the available FPGA resources, except for the CNN AE, which uses up to 47\% of the look-up tables (LUTs).
Moreover, the latency is less than about 365\unit{ns} for all models except the CNN AE, which has a latency of 1480\,\unit{ns}.
The II for all models is within the required 115\unit{ns}, again except the CNN AE.
Based on these, both types of architectures with both types of autoencoders are suitable for application at the LHC L1T, except for the CNN AE, which consumes too much of the resources.

Since the performance of all the models under study are of a similar level, we choose the ``best" model based on the smallest resource consumption, which turns out to be DNN VAE. 
This model was integrated into the $\texttt{emp-fwk}$ infrastructure firmware for LHC trigger boards~\cite{emp-fwk}, targeting a Xilinx VCU118 development kit, with the same VU9P FPGA as previously discussed.
Data were loaded into onboard buffers mimicking the manner in which data arrives from optical fibres in the L1T system.
The design was operated at $\SI{240}{MHz}$, and the model predictions observed at the output were consistent with those captured from the HLS C Simulation.
For this model we also provide resource and latency estimates for a Xilinx Virtex 7 690 FPGA, which is the FPGA most widely used in the current CMS trigger. The estimates are given in Table~\ref{tab:resources_v7}.

\begin{table*}[tb]
    \centering
    \caption{Resource utilization and latency for the quantized and pruned DNN and CNN (V)AE models.
    Resources are based on the Vivado estimates from Vivado HLS 2020.1 for a clock period of 5\unit{ns} on Xilinx VU9P.
    \label{tab:resources}}
    \begin{tabular}{l|c|c|c|c|c|c|c}
        \hline
        Model  & DSP~[\%] & LUT~[\%] & FF~[\%] & BRAM~[\%] & Latency~[ns] & II [ns]\\
        \hline
        DNN AE QAT 8 bits & 2 & 5 & 1 & 0.5 & 130 & 5 \\
        \hline
        CNN AE QAT 4 bits  & 8 & 47 & 5 & 6 & 1480 & 895 \\
        \hline
        \hline
        DNN VAE PTQ 8 bits & 1 & 3 & 0.5 & 0.3 & 80 & 5 \\
        \hline
        CNN VAE PTQ 8 bits & 10 & 12 & 4 & 2 & 365 & 115 \\
        \hline
    \end{tabular}
\end{table*}

\begin{table*}[tb]
    \centering
    \caption{Resource utilization and latency for the quantized and pruned DNN AE model.
    Resources are based on the Vivado estimates from Vivado HLS 2020.1 for a clock period of 5\unit{ns} on Xilinx V7-690.
    \label{tab:resources_v7}}
    \begin{tabular}{l|c|c|c|c|c|c|c}
        \hline
        Model  & DSP~[\%] & LUT~[\%] & FF~[\%] & BRAM~[\%] & Latency~[ns] & II [ns]\\
        \hline
        DNN VAE PTQ 8 bits & 3 & 9 & 3 & 0.4 & 205 & 5 \\
        \hline
    \end{tabular}
\end{table*}

\section{Conclusions}
\label{sec:conclusion}
We discussed how to extend new physics detection strategies at the LHC with autoencoders deployed in the L1T infrastructure of the experiments.
In particular, we show how one could deploy a deep neural network (DNN) or convolutional neural network (CNN) AE on a field-programmable gate array (FPGA) using the \hlsfml library, 
within a $\mathcal{O}(1)\mu\mathrm{s}$ latency and with small resource utilization once the model is quantized and pruned. 
We show that one can retain accuracy by compressing the model at training time. 
Moreover, we discuss different strategies to identify potential anomalies. 
We show that one could perform the anomaly detection (AD) with a variational AE (VAE) using the projected representation of a given input in the latent space, which has several advantages for an FPGA implementation: (1) no need to sample Gaussian-distributed pseudorandom numbers
(preserving the deterministic outcome of the trigger decision) and (2) no need to run the decoder in the trigger, resulting in a significant resource saving. 

As can be seen from Table~\ref{tab:resources}, the latency when using only the encoder as opposed to full VAE is reduced by a factor of two, while the performance is of a similar level (see Table~\ref{tab:performance_quantised}).
The DNN (V)AE models use less than 5\% of the Xilinx VU9P resources and the corresponding latency is within 130\unit{ns}, 
while the CNN VAE uses less than 12\% and the corresponding latency is 365\unit{ns}. 
All three models have the initiation interval within the strict limit imposed by the frequency of bunch crossing at the LHC.
Between the two architectures under study, the DNN requires a few times less resources in the trigger, however both DNN and CNN fit the strict latency requirement and therefore both architectures can potentially be used at the LHC trigger.
The CNN AE model is found to require more resources than are available.

With this work, we have identified and finalized the necessary ingredients to deploy (V)AEs in the L1T of the LHC experiments for Run 3 to accelerate the search for unexpected signatures of new physics.

\section{Code availability}
\label{sec:code}
The QKeras library is available under \url{github.com/google/qkeras}, where the work presented here is using QKeras version 0.9.0. 
The \hlsfml library with custom layers used in the paper are under \texttt{AE\_L1\_paper} branch and is available at \url{https://github.com/fastmachinelearning/hls4ml/tree/AE_L1_paper}. 

\section{Data availability}
\label{sec:data_availability}
The data used in this study are openly available at Zenodo at Ref.~\cite{zenodo_bkg_training,zenodo_AZZ,zenodo_LQ, zenodo_htautau, zenodo_htaunu}.

\section{Author information}
\label{sec:author}
Correspondence and material requests can be e-mailed to E. Govorkova (katya.govorkova@cern.ch).

\section*{Acknowledgements}
This work is supported by the European Research Council (ERC) under the European Union's Horizon 2020 research and innovation program (Grant Agreement No. 772369) and the ERC-POC programme (grant No. 996696).

\newpage

\bibliographystyle{naturemag}
\bibliography{biblio}

\end{document}